\documentstyle[epsfig,12pt]{article}
\catcode`\@=11

\@addtoreset{equation}{section}

\def\blankcite{\@ifnextchar
[{\@tempswatrue\@blankcitex}{\@tempswafalse\@blankcitex[]}}
\def\@blankcitex[#1]#2{%
  \let\@citea\@empty
  \@blankcite{\@for\@citeb:=#2\do
    {\@citea\def\@citea{,\penalty\@m\ }%
     \edef\@citeb{\expandafter\@iden\@citeb}%
     \if@filesw\immediate\write\@auxout{\string\citation{\@citeb}}\fi
     \@ifundefined{b@\@citeb}{{\reset@font\bfseries ?}%
       \G@refundefinedtrue\@latex@warning
       {Citation `\@citeb' on page \thepage \space undefined}}%
     {\hbox{\csname b@\@citeb\endcsname}}}}{#1}}
\def\@blankcite#1#2{{#1\if@tempswa , #2\fi}}

\catcode`@=12

\newfont{\mybb}{msbm10 scaled 1200}

\newcommand{\ind}{\hspace{.5cm}}

\newcommand{\be}{\begin{equation}}
\newcommand{\ee}{\end{equation}}
\newcommand{\bea}{\begin{eqnarray}}
\newcommand{\eea}{\end{eqnarray}}
\newcommand{\m}{{\scriptscriptstyle -}}
\newcommand{\p}{{\scriptscriptstyle +}}
\newcommand{\sfrac}[2]{{\textstyle \frac{#1}{#2}}}
\newcommand{\intl}{\int\limits_{-L}^L}
\newcommand{\sgn}{{\mbox{sgn}}}
\newcommand{\om}{\omega}

\newcommand{\vi}{\varphi}
\newcommand{\bra}{\langle}
\newcommand{\ket}{\rangle}
\newcommand{\la}{\lambda}
\newcommand{\fa}{\frac{1}{2L}}
\newcommand{\fb}{\frac{1}{(2L)^2}}

\newcommand{\Proj}{\mbox{\mybb P}}
\newcommand{\Pn}{\Proj_{\mbox{\scriptsize \sf N}}}
\newcommand{\sfn}{\mbox{\scriptsize \sf N}}
\newcommand{\Pnm}{\Proj_{\mbox{\scriptsize \sf N-1}}}

\renewcommand{\thefootnote}{\alph{footnote}}

\begin{document}

\thispagestyle{empty}

\hfill hep-th/9512179, TPR-95-20
\vspace{1cm}

\begin{center}
{\huge The Vacuum Structure of Light-Front $\phi^4_{1+1}$-Theory}

\vspace{1.5cm}

{\large
T.~Heinzl\footnote{e-mail: thomas.heinzl@physik.uni-regensburg.de},
C.~Stern\footnote{e-mail: christian.stern@physik.uni-regensburg.de},
E.~Werner\footnote{e-mail:   ernst.werner@physik.uni-regensburg.de} \\
Universit\"at  Regensburg, Institut f\"ur Theoretische Physik, \\
93053 Regensburg, FRG

\vspace{1cm}

B.~Zellermann\footnote{e-mail:   bernd@ptprs6.phy.tu-dresden.de},
\\
TU     Dresden,     Fakult\"at      f\"ur     Mathematik      und
Naturwissenschaften,\\
Institut  f\"ur Theoretische  Physik,  \\
01062 Dresden, FRG}

\end{center}

\vspace{1.5cm}

\begin{abstract}
We  discuss  the  vacuum  structure  of  $\phi^4$-theory  in 1+1
dimensions  quantised  on the light-front $x^+ =0$.  To this end,
one  has  to  solve  a  non-linear,  operator-valued   constraint
equation.   It expresses  that mode of the field operator  having
longitudinal light-front momentum equal to zero, as a function of
all the other modes in the theory.   We analyse whether this zero
mode can lead to a non-vanishing  vacuum expectation value of the
field  $\phi$  and thus  to spontaneous  symmetry  breaking.   In
perturbation theory, we get no symmetry breaking. If we solve the
constraint, however, non-perturbatively, within a mean-field type
Fock ansatz, the situation changes: while the vacuum state itself
remains trivial, we find a non-vanishing vacuum expectation value
above a critical  coupling.   Exactly the same result is obtained
within   a  light-front   Tamm-Dancoff   approximation,   if  the
renormalisation is done in the correct way.
\end{abstract}

\vfill
\newpage

\renewcommand{\thefootnote}{\arabic{footnote}}
\setcounter{footnote}{0}
\setcounter{page}{1}

\section{Introduction}

\vspace{1cm}

Back in 1949, Dirac \cite{Dir49} noted that within a relativistic
formulation  of  Hamiltonian  dynamics  the  choice  of the  time
evolution  parameter $\tau$ is not unique.  As an alternative  to
the usual  cartesian  time $\tau  = t$ of Galileian  dynamics  he
suggested  the variable  $\tau = x^\p \equiv t + z/c$, now called
``light-cone  time".  For this choice, hyperplanes tangent to the
light-cone,  {\it  i.e.}~light-fronts  or  null-planes,  $x^\p  =
const$, are surfaces  of equal time.  The associated  Hamiltonian
$H$  then  describes  the  time  evolution  in $x^\p$  off  these
surfaces.

\ind
Independently,  in the  early  sixties,  the  idea  of specifying
initial data on null-hyper\-sur\-faces  was developed  by Penrose
and others  with  particular  regard  to the gravitational  field
\cite{Pen60}.

\ind
It was  not before  the  late  sixties,  however,  when  particle
physicists became aware of Dirac's work. It was realized that the
infinite-momentum  limit  used  in  current  algebra  \cite{FF65}
amounts to using ``light-like"  charges defined as integrals over
null-planes   \cite{BS67}.     Soon   after   this   observation,
quantisation  of field  theories  on light-fronts  was explicitly
formulated \cite{KS70,LKS70,NR71}.   The first reference to Dirac
was made in a paper  by Chang  and Ma \cite{CM69}  on light-front
(LF) perturbation  theory.  For reviews on the early achievements
of LF field theory see \cite{AFF73}.

\ind
Already  at that time it was repeatedly  stated  that  one of the
main  advantages  of  the  infinite-momentum   frame  (and  hence
light-front  quantisation)   is  the  simplicity  of  the  vacuum
structure.   As early  as 1966,  Weinberg  realized  that  within
``old-fashioned"     Hamiltonian    perturbation    theory    the
infinite-momentum  limit of many diagrams,  in particular  vacuum
diagrams,  is vanishing \cite{Wei66}.   The success of this limit
for current algebra  was traced in \cite{LKS70}  to the fact that
light-like charges always annihilate the vacuum, irrespective  of
whether  they  are  conserved  or not. Thus,   Coleman's  theorem
\cite{Col66}, ``the symmetry of the vacuum is the symmetry of the
world", does not apply.  Furthermore,  the Fock-vacuum,  {\it i.e.}~the
ground   state   of  the  free  Hamiltonian,   is  stable   under
interaction.   In the same manner as the light-like  charges, the
fully interacting Hamiltonian annihilates the vacuum, at least in
theories with a mass gap \cite{LKS70}.   The technical reason for
this is the positivity of a kinematical Poincar\'e generator, the
longitudinal  momentum  $P^\p  = P^0  + P^3$:   as any  (massive)
particle carries positive $p^\p$ it cannot be degenerate with the
vacuum  having  $p^\p = 0$.  These results  were then transformed
into the folkloristic  statement:  "on the light-cone, the vacuum
is trivial".

\ind
With  the  advent  of QCD as the theory  of strong  interactions,
however, people began to feel uneasy with this statement.   There
was (and is) growing evidence,  that many of the phenomenological
aspects of hadron physics,  like confinement  and chiral symmetry
breaking  are  related  to the non-trivial  features  of the  QCD
vacuum (within standard  equal-time quantisation  on a space-like
hypersurface). Let us only mention features like quark- and gluon
condensates,  instantons,  monopole condensation  etc., which all
indicate  that  the vacuum  is densely  populated  by non-trivial
quantum  fluctuations,  which furthermore  are not accessible  to
perturbation theory.

\ind
The  concern  that  arose  at these  points  can be put into  the
question: can the existence of these large vacuum fluctuations be
reconciled  with the triviality  of the light-front  vacuum?  For
QCD, the answer to this question is not (yet) known.  For simpler
theories, it depends to some extent on the theory. Generally, one
can say the following.   Not unexpectedly,  the delicate point is
the behaviour of the degrees of freedom at longitudinal  momentum
$p^\p = 0$. At this point, even the energy\footnote{our
LF conventions  are:   $x^\pm  = (x^0  \pm
x^3)/\sqrt{2}$,  $ \partial_\pm  = \partial  / \partial{x^\pm}$.}

\be
p^- = \frac{p_\perp^2 + m^2}{2p^\p}
\ee

of a free particle of mass $m$ diverges.  In perturbation theory,
one  encounters  associated  infrared  divergences.    These  are
conveniently  regularised  by working in a finite spatial volume.
In this way, the modes  having  $p^\p  =0$, shortly  called  zero
modes     (ZMs),     can    be    explicitly     isolated     and
studied [\blankcite{MY76} -- \blankcite{WX95}].  It is generally believed that
these  modes  carry  the  information  on the non-trivial  vacuum
aspects.   This  has been  shown  in particular  for $\phi^4$-theory
in $d=1+1$, where the ZMs are responsible  for spontaneous
symmetry breaking \cite{HKS92,  Rob93, BPS93, PS94, HPS95}.  This
article  elaborates  on the quoted works on $\phi^4_{1+1}$-theory
by  extending  the  approximations   made  there  and  clarifying
subtleties  of  the  renormalisation  procedure.   The  different
approaches are compared in detail.

\ind
This  paper  is organized  as follows.   In Section  2 we shortly
review the canonical  phase space structure  of LF field theories
which generically display constraints.  For LF $\phi^4_{1+1}$, it
is the vacuum  expectation  value  of the field  $\phi$  that  is
constrained.    This  constraint  is  solved  by  a  perturbative
expansion  in Section 3, by a mean-field Fock ansatz in Section 4
and within a LF Tamm-Dancoff approximation in Section 5.

\vspace{1cm}
\newpage

\section{Light-Front Field Theory as a First Order System}

\vspace{1cm}

The Lagrangian of LF scalar field theory in 1+1 dimensions

\be
L[\phi,  \partial_\p  \phi]  = \int  dx^{\m}  \partial_{\m}  \phi
\partial_\p \phi - U [\phi] \label{LFLag}
\ee

is   {\it   linear}   ({\it   i.e.}~first   order)   in  the   LF
velocity
$\partial_{\p}  \phi$.  The quantisation  of such systems  is not
quite straightforward. People commonly refer to Dirac's treatment
of constrained dynamics \cite{Dir50, HRT76} with its categorizing
of constraints  into primary,  secondary,  ...,  first-class  and
second-class.   For LF field theories this was first employed  by
Banyai, Mezinescu \cite{BM73} and others \cite{MY76,HRT76,Ida76}.
There is, however,  a much more economic  method  for first-order
systems due to Faddeev and Jackiw \cite{FJ88},  which was to some
extent anticipated  in \cite{Gar71}.   In the context of LF field
theory  it  has  only  been  used  in a few  recent  publications
\cite{HZ93a,  HW94}.   We shortly  review the method for a finite
number  of degrees  of freedom  and concentrate  on those  issues
which will be relevant for LF field theory.

Consider a Lagrangian

\be
L (x) = \frac{1}{2}x_i  f_{ij}\dot  x_j - \Phi(x)\; ; \quad i,j =
1, \ldots N \; ,
\label{FJL}
\ee

where the matrix $f_{ij}$ is antisymmetric. If it has an inverse,
$f_{ij}^{-1}$,  the canonical  bracket between  the $x$-variables
(generalizing the well-known Poisson bracket) is

\be
\{x_i , x_j\} = f_{ij}^{-1} \; .
\ee

In this case the number  $N$ of $x$'s  must be even,  and one can
introduce   canonical   coordinates    $q_\alpha$   and   momenta
$\pi_\alpha$   ({\it i.e.}~a  polarization  and  Poisson  brackets)  as
discussed extensively in Ref.~\cite{CN94}. If the matrix $f_{ij}$
does not have an inverse, there are zero modes $z^a$, satisfying

\be
f_{ij} z_j^a = 0 \; .
\ee

The Lagrangian (\ref{FJL}) can then be cast into the form

\be
L (y,z) = \frac{1}{2}  y_m \hat f_{mn}  \dot y_n - H (y, z) \; ;
\quad m,n = 1, \ldots N' \; ,
\label{FJL2}
\ee

where  the matrix  $\hat f_{mn}$  is the invertible  sub-block  of
$f_{ij}$, thus $N'< N$ and the number $N'$ of $y$'s is even.  $H$
denotes the Hamiltonian.   The $z$-variables  are constrained via
their equation of motion,

\be
\frac{\partial H}{\partial z^a} = 0 \; ,
\ee

and, as stressed  by Faddeev and Jackiw,  these are the only true
constraints  in the theory.  They should be used to eliminate the
$z$'s,    which    might   turn   out   to   be   difficult    or
impossible\footnote{The  particular  case  when $H$ is linear  in
(some  of) the $z$'s will not be considered  here.   It leads  to
additional constraints  between the $y$-variables  and is typical
for gauge theories, where these constraints correspond to Gauss's
law\cite{FJ88}.}.   The  $y$'s  are  the  unconstrained,  ``true"
degrees of freedom. They have the canonical bracket

\be
\{y_m , y_n \} = \hat f^{-1}_{mn}  \; ; \quad m,n = 1, \ldots  N'
\; .
\ee

\vspace{.5cm}

Let us extend  this  discussion  to field  theory.   The main new
problem  arising is of course the fact that the number of degrees
of  freedom  becomes   infinite.    Matrices   therefore   become
differential   operators.    If  one  looks   at  the  Lagrangian
(\ref{LFLag}),  one readily  notes  that the matrix  $f_{ij}$  is
replaced   by   the   spatial   derivative    $\partial_{\m}    =
\partial/\partial  x^{\m}$.  In  order  to  uniquely  define  its
inverse one has to specify its domain and boundary conditions. We
therefore enclose our spatial variable $x^{\m}$ in a box, $-L \le
x^{\m} \le L$, and impose periodic boundary  conditions  (pBC) on
our  fields.   This  is,  of  course,  nothing  but  an  infrared
regularisation.   In this case, the operator $\partial_{\m}$  has
zero modes, namely all spatially constant functions.

If  we  split  our  field  $\phi$  into  a ZM  $\omega$  and  its
complement $\varphi$,

\bea
\label{OMVI}
\phi (x^{\p} , x^{\m}) &=& \omega (x^{\p}) + \varphi  (x^{\p} , x^{\m} ) \;
, \\
\omega (x^{\p})  &\equiv&  \frac{1}{2L}  \intl dx^{\m} \phi (x^{\p} ,
x^{\m}) \; ,
\eea

such that

\be
\intl dx^{\m} \varphi (x^{\p} , x^{\m} ) = 0 \; ,
\ee

the  Lagrangian   (\ref{LFLag})   can  be  rewritten  analogous   to
(\ref{FJL2}),

\be
L[\varphi,   \omega]   =  \intl   dx^{\m}   \frac{1}{2}   \varphi
(-2\partial_{\m}) \dot \varphi - H [\varphi, \omega] \; .
\ee

The  Hamiltonian  $H$ is identical  with  the potential  $U$ from
(\ref{LFLag})   after  the  replacement  (\ref{OMVI})   has  been
performed.   Thus we see that the ZM $\omega$  is the analogue of
the $z$-variables and therefore constrained via

\be
\frac{\delta H}{\delta \omega} = 0 \; .
\label{ZMC}
\ee

The basic bracket is given by

\be
\{\varphi  (x^{\p} , x^{\m} ) , \varphi (x^{\p} , y^{\m} ) \} = -
\frac{1}{2}\langle  x^\m \vert  \partial_{\m}^{-1}  \vert  y^{\m}
\rangle \; , \label{COMMPHI}
\ee

where the matrix element on the {\it r.h.s.}~denotes  the periodic sign
function\cite{HW94}

\be
\langle  x^{\m}  \vert  \partial_{\m}^{-1}  \vert y^{\m}  \rangle
\equiv \frac{1}{2} \sgn (x^{\m} - y^{\m} ) - \frac{x^{\m} - y^{\m}
}{2L} \; .
\ee

The discussion above becomes especially  transparent  if one goes
to momentum space. Expanding the field into Fourier modes

\be
\phi = a_0 + \sum_{n=1}^\infty  \frac{1}{\sqrt{4\pi  n}} (a_n
e^{-i \pi n x^\m /L} + a_n^* e^{i \pi n x^\m /L} ) \; ,
\label{FOURIER}
\ee

the  Lagrangian  becomes  (discarding  a  total  time  derivative
\cite{FJ88})

\be
L(a_n,  a_0) = -i \sum_{n>0} a_n \dot a_n^*  - H(a_n , a_0) \; ,
\ee

where $a_0 \equiv \omega$ is the constrained ZM.  If we introduce
a momentum cutoff $N$, $n <N$, we have mapped the field theory on
a finite dimensional  system.  The elementary bracket between the
Fourier modes can be read off from the kinetic term,

\be
\{a_m , a_n^* \} = -i \delta_{mn} \; .
\label{ELBRACK}
\ee

Quantisation   is  performed  by  employing   the  correspondence
principle,  {\it i.e.}~by replacing $i$ times the canonical  bracket by
the commutator.   For arbitrary classical observables,  $A$, $B$,
this means

\be
[\hat A, \hat B] = i \widehat{\{A, B\}} \; ,
\ee

so that, from (\ref{ELBRACK}), our elementary commutator becomes

\be
[a_m , a_n^\dagger ] = \delta_{mn} \; .
\ee

As is well known, not all classical  observables  $A$, $B$ can be
quantised   unambiguously   due  to  possible  operator  ordering
problems.  Such problems do not arise for the field $\varphi$ and
the bracket  (\ref{COMMPHI}),  where  the field-independent  {\it
r.h.s.}~leads   to  a  $c$-number  commutator.    The  constraint
(\ref{ZMC}),  however, implies a functional dependence  of the ZM
$\omega$  on $\varphi$  and thus  a non-vanishing  commutator  of
$\om$ with $\vi$.  This can be explicitly verified by calculating
the associated Dirac bracket within the Dirac-Bergmann  algorithm
\cite{HKW91b, HWZ92}.  For the quantum theory, this results in an
ordering ambiguity with respect to $\om$ and $\vi$.  Therefore, a
definite  ordering  has  to  be prescribed.   We chose  Weyl  (or
symmetric)  ordering  \cite{HKS92,   PS94}  which  is  explicitly
hermitian.   Using  this  prescription,  one  finds  the  quantum
Hamiltonian for LF $\phi^4_{1+1}$-theory,

\bea
H   =   \intl   dx^\m \!\!\! &\Bigg(& \!\!\!\! \frac{1}{2} m^2 \vi^2 +
\frac{\lambda}{4!} \vi^4\Bigg) + \nonumber \\
+   \intl    dx^\m  \!\!\!\! &\Bigg[& \!\!\!\! \frac{1}{2}    m^2   \om^2    +
\frac{\lambda}{4!}  \Big( \om^4  + \om\vi^3  + \vi \om \vi^2  +
\vi^2 \om \vi + \vi^3 \om + \nonumber \\
&& + \, \om^2 \vi^2 + \vi^2 \om^2 + \om \vi \om \vi + \vi \om \vi \om
+ \om \vi^2 \om + \vi \om^2 \vi \Big) \Bigg] \; .
\label{HAM}
\eea

Note that we have chosen  the sign of the mass term(s)  in such a
way that there is no spontaneous symmetry breaking at tree level.
With this Hamiltonian,  equation (\ref{ZMC})  for the constrained
ZM $\omega$ reads explicitly

\be
\theta = \frac{\delta H}{\delta \om}   \equiv    m^2    \om   +
\frac{\lambda}{3!}\om^3    +
\frac{\lambda}{3!}  \frac{1}{2L} \intl dx^\m \left [\vi^3 + \vi^2
\om + \vi \om \vi + \om \vi^2 \right] = 0 \; . \label{T1}
\ee

This  is nothing  but the ZM of the  Euler-Lagrange  equation  of
motion for the total field $\phi$ decomposed into $\om$ and $\vi$
\cite{MY76,  Wit89,  PS94}.   The  remainder  of  this  paper  is
concerned  with different  approaches  to solve this equation for
$\omega$.

\vspace{1cm}

\section{Perturbative Solution}

\vspace{1cm}

To obtain  a perturbative  solution  for $\om$ we expand  it in a
power series in $\lambda$,

\be
\om  \equiv \sum_{n=0}^{\infty} \la^{n}\om_{n}
\label{SERIES}
\ee

Inserting  this  into  (\ref{T1})  determines   the  coefficients
$\om_n$ recursively. For the first three we find

\bea
\om_0 &=& 0 \; , \\
\om_{1}   &=&   -\frac{1}{6m^2}\fa\intl    dx^\m   \vi^3(x)\,   ,
\label{OM1} \\
\om_{2}   &=&  \frac{1}{36m^4}   \fb  \intl  dx^\m  dy^\m  \left[
\vi^2(x)\vi^3(y) + \vi(x)\vi^3(y)\vi(x) + \vi^3(y)\vi^2(x)\right]
\; .
\eea

All  higher  orders  may be obtained  similarly.   Unfortunately,
however,  we have  not been able  to find  a closed  formula  for
$\om_n$ in order to sum up the whole series (\ref{SERIES}).

If   we   expand   the   quantum   field   $\vi$   according   to
(\ref{FOURIER}) in terms of Fock operators,

\be
\vi(x) = \sum_{n=1}^{\infty}  \frac{1}{\sqrt{4\pi  n}}\Big [a_{n}
e^{-ik_{n}^{+} x^{-}} + a_{n}^{\dagger} e^{ik_{n}^{+}x^{-}} \Big]
\; ,
\label{FOCK}
\ee

with the discretised  longitudinal  momentum  $k^\p_n = \pi n/L$,
one notes  that the vacuum  expectation  value  (VEV) of $\om$ is
zero to all orders in $\la$, since $\om_n$ contains  an {\it odd}
number of Fock operators, {\it i.e.}~$\bra 0 \vert \om_n \vert 0 \ket =
0$. Thus

\be
\bra 0 \vert \om \vert  0 \ket = \sum_{n=0}^\infty  \la^n  \bra 0
\vert \om_n \vert 0 \ket = 0 \; .
\ee

This seems to imply  that a non-vanishing  VEV for $\om$ can only
arise non-perturbatively and must be non-analytic in the coupling
$\la$.   We will  discuss  this issue  in the next  sections  and
continue for the time being within the framework  of perturbation
theory.  In particular,  we will study the effect of the ZM $\om$
on the mass renormalisation.

To this  end, we split  up the Hamiltonian  $H$ from  (\ref{HAM})
into two pieces,

\be
H = H_0 + H_\om \; ,
\ee

where $H_0$ is independent of $\om$,

\be
\label{HZERO}
H_0 = \intl dx^\m \left(  \frac{1}{2}  m^2 \vi^2 + \frac{\la}{4!}
\vi^4 \right) \; ,
\ee

and $H_\om$ is the $\om$-dependent interaction,

\bea
H_\om   \equiv \intl dx^\m  \!\!\!\! &\Bigg[& \!\!\!\! \frac{1}{2} m^2 \om^2 +
\frac{\lambda}{4!}  \Big( \om^4  + \om\vi^3  + \vi \om \vi^2  +
\vi^2 \om \vi + \vi^3 \om + \nonumber \\
&& + \, \om^2 \vi^2 + \vi^2 \om^2 + \om \vi \om \vi + \vi \om \vi \om
+ \om \vi^2 \om + \vi \om^2 \vi \Big) \Bigg] \; .
\eea

This can be simplified using the constraint  equation  (\ref{T1})

\bea
H_\om &=& H_\om - \frac{L}{2}\left(\om \, \theta + \theta \, \om \right)
\nonumber \\
&=& \frac{\la}{4!} \intl  dx^\m
\left( -\om^4 + \vi \om \vi^2 + \vi^2 \om \vi
+ \vi \om^2 \vi - \om \vi^2 \om \right) \,.
\label{HOM}
\eea

Because $\om$  is of order  $\la$,  the ZM dependent  part
$H_\om$  is of order $\la^2$. Explicitly, we find

\bea
H_\om  &=& \frac{\la}{4!}  \intl dx^\m \left ( \vi \om \vi^2 + \vi^2
\om \vi \right) + O (\la^3)  \nonumber\\
&=& - \frac{\la^2}{144m^2} \fa \intl dx^\m dy^\m \left[ \vi(x) \vi^3(y)
\vi^2(x) + \vi^2(x) \vi^3(y) \vi(x) \right] + O (\la^3) \, ,
\label{CORR}
\eea

where we have used the first order term (\ref{OM1}). This induces
a mass shift of order $\la^2$ which is given by \cite{HWZ92} (see
also  \cite{Rob93})

\be
\delta  m^2 \equiv \frac{2\pi  n}{L} \Big( \bra n \vert H_\om
\vert n \ket - \bra 0 \vert H_\om \vert 0 \ket \Big) \; .
\label{SHIFT}
\ee

Here, $\vert n \ket \equiv a_n^\dagger \vert 0 \ket$ denotes a
one-particle state of longitudinal momentum $k^\p_n = \pi n/L$. We
have subtracted the constant vacuum energy, $\bra 0 \vert H_\om
\vert 0 \ket$, which diverges linearly with the volume. This is
sufficient to render the mass shift finite.  After inserting the
Fock-expansion (\ref{FOCK}) into (\ref{CORR}) one obtains for
(\ref{SHIFT})

\bea
\delta m^2 &=& -\frac{\la^2}{6m^2} \frac{L}{(4\pi)^3} \frac{1}{n}
\bigg[ \sum_{m=1}^{n-1} \frac{1}{(n-m)m} +
4 \sum_{m=1}^{\infty} \frac{1} {(n+m)m} \bigg]
\frac{2\pi n}{L} \nonumber\\
&=& -\frac{\la^2}{6m^2} \frac{L}{(4\pi)^3} \frac{2}{n^2} \left[
3\gamma + \Psi(n) + 2\Psi(1+n) \right] \frac{2\pi n}{L}  < 0 \;,
\label{SHIFTRES}
\eea

where  $\gamma$  is  Euler's  constant  and  $\Psi$  denotes  the
Digamma-function  \cite{AS70}.   The question now is, whether the
result   (\ref{SHIFTRES})   is  a  finite   size   effect,   {\it
i.e.}~vanishing  in the infinite  volume  limit.   The latter  is
obtained by replacing

\be
\sum_{n=1}^N   f(n)   \to   \lim_{L   \to \infty}   \frac{L}{\pi}
\int\limits_{\pi/L}^{\pi N/L} dk^\p f(k^\p ) \; .
\ee

in such  a way that  $k^\p_n$  approaches  a {\it  finite}  limit
$k^\p$.   For the case  at hand, this  amounts  to replacing  the
Digamma-function by its asymptotics \cite{AS70}

\be
\Psi(z) \simeq \mbox{ln}(z) - \frac{1}{2z} -\frac{1}{12z^2} +
\frac{1}{120z^4} + \cdots  \quad (z \rightarrow \infty \quad
\mbox{in} \quad \vert\mbox{arg}(z) \vert  < \pi)  \,.
\ee

Thus we finally obtain

\be
\lim_{L\to\infty} \delta m^2 = -\frac{\la^2}{m^2}  \frac{1}{16\pi
k^{\p}}    \lim_{L\to\infty}     \frac{1}{L}    \bigg[\gamma    +
3 \, \mbox{ln}(k^{\p} L/\pi) + \cdots \bigg] = 0 \,.
\ee

The vanishing  of this  expression  implies  that  the ZM induced
second  order  mass shift $\delta  m^2$ is indeed a finite  size
effect. We do not have a general proof that this is also true for
higher orders in $\la$.  However, it seems to be plausible that a
{\it single} mode like $\om$ constitutes a ``set of measure zero"
within  the  infinite  number  of modes  as long  as one  applies
perturbation theory. From the theory of condensation, however, it
is well known that single modes, especially  those with vanishing
momentum, can significantly  alter the perturbative  results.  To
analyse this possibility  one clearly has to use non-perturbative
methods. This will done in the next sections.

\vspace{1cm}

\section{Mean-Field Ansatz}

\vspace{1cm}

As $\phi^4_{1+1}$-theory  is super-renormalisable,  the number of
divergent  diagrams is finite, namely one:  the tadpole resulting
from a self-contraction of the field at the same space-time point.
Therefore,  renormalisation  can,  at least  within  perturbation
theory, be done via normal-ordering,  which is nothing but making
use of Wick's theorem:  expanding  the appearing powers of $\phi$
in  a  sum   of  normal-ordered   terms   with   more   and  more
self-contractions,  one separates  the convergent  term (with  no
contraction)   from  the  divergent   ones  (with  at  least  one
contraction).  The latter terms then are just the negative of the
required counterterms. For conventional $\phi^4$-theory one finds

\bea
\frac{1}{2}m^2 \phi^2 + \frac{\la}{4!} \phi^4 &=&
\frac{1}{2}m^2 \Big( :\! \phi^2 \!\! : + \, T \Big) +
\frac{\la}{4!} \Big( :\! \phi^4 \!\! : +\, 6\, T :\! \phi^2 \!\! : +\, 3\, T^2
\Big) \nonumber \\
&=& \frac{1}{2} \left( m^2 + \frac{\la}{2} T\right) :\! \phi^2 \!\! : +\,
\frac{\la}{4!} :\! \phi^4 \!\! : +\, \frac{m^2}{2} T + \frac{\la}{8} \, T^2 \;
,
\label{WICK}
\eea

where

\be
T   \equiv    \bra   \phi^2    \ket   =   \int    \frac{dk}{4\pi}
\frac{1}{\sqrt{k^2 + m^2}}
\label{ET-TADPOLE} \ee

formally   denotes   the   logarithmically    divergent   tadpole
contribution  (in  $d$ = 1+1)  which  coincides  with  the VEV of
$\phi^2$ or a self-contraction of the field.

\ind
It is obvious  that only mass and vacuum energy get renormalised.
The  renormalised   Hamiltonian   is  obtained   by  adding   the
counterterms

\be
\delta{\cal H} \equiv - \frac{\la}{4} T : \phi^2 :
- \frac{1}{2}m^2  T  - \frac{\la}{8}  T^2 \; ,
\label{COUNTER}
\ee

If the VEV of $\phi^2$ is taken in a Fock vacuum corresponding to
the bare  mass $m$, the latter  coincides  with  the renormalised
mass to order $\la$.

\ind
This  rather  trivial   renormalisation   procedure   cannot   be
straightforwardly  extended to LF field theory, simply because we
do not  know  the  Hamiltonian!   The  ZM $\om$  is a complicated
functional of the Fock operators $a_n, a_n^\dagger$, which has to
be  found   from  the  constraint   (\ref{T1})   before   we  can
normal-order. There is, however, a way around this obstacle, such
that  an  {\it  exact}  knowledge  of  $\om$  is not  needed  for
renormalisation.

\ind
To this end we use the following general ansatz for $\om$:

\bea
\om &=& \om_0  + \sum_{n>0}  \om_n  a_{n}^\dagger  a_n +
\sum_{m,n >0} \om_{mn} a_{m}^\dagger a_{n}^\dagger a_{m+n} +
\sum_{m,n>0} \om_{mn}^* a_{m+n}^\dagger a_m a_n + \nonumber \\
&+&  \sum_{l,m,n>0}  \om_{lmn} \, a^\dagger_{l+m+n}  a_l  a_m a_n +
\sum_{l,m,n>0}  \om_{lmn}^* \, a_l^\dagger  a_m^\dagger a_n^\dagger
a_{l+m+n} + \nonumber \\
&+& \sum_{k,l,m,n>0}  \delta_{k+l,  m+n} \,  \om_{klmn} \, a_k^\dagger
a_l^\dagger a_m a_n + \dots  \; .
\label{MFA}
\eea

This ansatz  is hermitian  and takes  care of the fact that $\om$
cannot  transfer  any momentum.   It can be understood  as a Wick
expansion  written in the opposite of the usual order:  the first
term $\om_0$  is the sum of all contractions,  the second the sum
of  all  contractions  but  one  and  so  on.  Accordingly,  each
individual term in the expansion is a normal-ordered operator. In
\cite{HKS92}  this ansatz  was used (with a truncation  after the
second term) to determine  $\om$ and the vacuum structure  of the
theory.   In the following  we will  analyse  the renormalisation
structure   in  more  detail  and  extend  the  analysis  to  the
calculation  of one-particle  energies  and the  ZM induced  mass
shift.

\ind
Inserting this ansatz into (\ref{HAM})  and (\ref{T1})  we obtain
for the constraint and the Hamiltonian

\bea
\theta  &=&  \theta_0  + \sum_{n>0}  \theta_n  a_n^\dagger  a_n +
\ldots \; ,
\label{MFT} \\
H &=& H_0 + \sum_{n>0} H_n a_n^\dagger a_n + \ldots \; ,
\label{MFHAM}
\eea

where,   in  accordance   with  the  truncation   of  our  ansatz
(\ref{MFA}),  we have omitted terms containing more than two Fock
operators.    The  coefficients   $H_0$,  $H_n$  and  $\theta_0$,
$\theta_n$   are  functions   of  $\om_0$  and  $\om_n$.    Thus,
(\ref{MFHAM})  is an effective one-body or mean-field (MF) Hamiltonian
describing the influence of the ZM $\om$.  Explicitly,  one finds
for the coefficients of the constraint

\bea
\label{T0MF}
\theta_0   &=&  \left(m^2   +  \frac{\la}{2}   T  \right)  \om_0  +
\frac{\la}{3!}  \left( \om_0^3 + \sum_{n>0} \frac{\om_n}{4\pi  n}
\right) \; , \\
\label{T2}
\theta_n  &=&  \left(  m^2  +  \frac{\la}{2}  T \right)  \om_n  +
\frac{\la}{3!} \left( \om_n^3 + 3 \om_0 \om_n^2 + 3 \om_0^2 \om_n
+ \frac{6\om_0}{4\pi  n}  + \frac{6\om_n}{4\pi  n}  \right)  \; ,
\eea

and of the Hamiltonian (scaled by 2$L$)

\bea
\label{E0MF}
\frac{H_0}{2L}  &=&  \frac{1}{2}\left(  m^2  +  \frac{\la}{2}  T  \right)
\om_0^2  +  \frac{\la}{4!}  \left(  \om_0^4  + 4  \sum_{n>0}
\frac{\om_0 \om_n}{4\pi n} + \sum_{n>0} \frac{\om_n^2}{4\pi  n} \right)
+ \frac{m^2}{2} T + \frac{\la}{8} T^2 , \\
\label{ENERGIES1}
\frac{H_n}{2L} &=& \frac{1}{2}\left( m^2 + \frac{\la}{2} T \right) \left(
\om_n^2 + 2\om_0 \om_n + \frac{1}{2\pi  n} \right) + \nonumber \\
&+&  \frac{\la}{4!}   \left(  (\om_0  +  \om_n)^4   -  \om_0^4  +
\frac{3}{\pi n} (\om_0 + \om_n)^2 + \frac{\om_n^2}{2\pi  n} \om_n
\sum_{m>0}  \frac{\om_m}{\pi  m} \right)  \; .
\eea

$T$  now denotes the LF tadpole (in discretised form)

\be
T= \bra \varphi^2 \ket = \sum_{n>0}\frac{1}{4\pi n} \; ,
\ee

which is mass independent in contrast to (\ref{ET-TADPOLE}).
Note that the one-particle  matrix elements  of $\theta$  and $H$
are given as the {\it sum} of two coefficients

\bea
\bra n |\theta | n \ket &=&  \theta_0 + \theta_n  = \nonumber \\
&=& \left( m^2 + \frac{\la}{2}  T \right) (\om_0 + \om_n)
+  \frac{\la}{3!}  \left[  (\om_0  + \om_n)^3  + 6 \frac{\om_0  +
\om_n}{4\pi  n} + \sum_{k>0}  \frac{\om_k}{4\pi  k} \right] ,
\label{TNMF}
\eea

and

\bea
\bra n | H/2L | n \ket &=& (H_0 + H_n)/2L = \nonumber \\
&=& \frac{1}{2} \left( m^2 + \frac{\la}{2} T \right) (\om_0 +
\om_n)^2  +  \frac{1}{2}  \left(  m^2  + \frac{\la}{2}  T \right)
\frac{1}{2\pi n} + \nonumber \\
&+&  \frac{\la}{4!}  \Bigg[  (\om_0  + \om_n)^4  + (12 \om_0^2  +
24\om_0 \om_n + 14 \om_n^2)  \frac{1}{4\pi  n} + \nonumber \\
&& \quad\quad +\: 4(\om_0 + \om_n )
\sum_{k>0} \frac{\om_k}{4\pi  k} + \sum_{k>0} \frac{\om_k^2}{4\pi
k} \Bigg] + \nonumber \\
&+& \frac{m^2}{2} T + \frac{\la}{8} T^2 \; .
\label{HNMF}
\eea

{}From  the  divergence  structure  above  it  is  clear  that  all
coefficients can be made finite by adding the counterterm

\be
\delta H/2L = - \frac{\la}{4}  T \sum_{n=1}^\infty  \frac{1}{2\pi n}
a_n^\dagger  a_n - \frac{\la}{4}  T \om^2  - \frac{1}{2}  m^2 T -
\frac{\la}{8} T^2 \; ,
\label{COUNTER2}
\ee

which can be obtained  from (\ref{COUNTER})  by integrating  over
$x^\m$  and decomposing  the field.  The renormalisation  is thus
standard, {\it i.e.}~performed by normal-ordering and formally achieved
by setting $T=0$ in the expressions above.

\ind
It is convenient to rescale the coefficients

\bea
\om_0 &\to& \frac{\om_0}{\sqrt{4\pi}} \; , \\
\om_n &\to& \frac{\om_n}{\sqrt{4\pi}} \; ,
\eea

and define a dimensionless coupling $g$ as

\be
g \equiv \frac{\la}{24\pi m^2} \; ,
\ee

such that (\ref{T0MF}) - (\ref{ENERGIES1}) become

\bea
H_0/2L &=& \frac{m^2}{4\pi} \Bigg[\frac{1}{2} \om_0^2+
   \frac{g}{4}\left( \om_0^4 + 4 \om_0 \sum_{n=1}^{\infty}
   {\frac{\om_n}{n}} +    \sum_{n=1}^{\infty} {\frac{\om_n^2}{n}}
   \right) \Bigg]  \; , \\
\label{ENERGIES2}
H_n/2L &=&  \frac{m^2}{4\pi} \Bigg[ \frac{1}{2}
    \left( \om_n^2 + 2\om_0 \om_n +\frac{2}{n} \right) + \nonumber  \\
    &+& \frac{g}{4}\left(  \left(  \om_0 +\om_n  \right)^4-\om_0^4  +
    \frac{12}{n}       \left(      \om_0+\om_n       \right)^2      +
    4\om_n\sum_{k=1}^{\infty}{\frac{\om_k}{k}}+2\frac{\om_n^2}{n}
    \right)
    \Bigg] \; ,
\eea

and

\bea
\label{ANSVAC}
\om_0 &+& g \left( \om_0^3 + \sum_{n=1}^{\infty} {\frac{\om_n}{n}}
\right) = 0  \; , \\
\label{ANSPART}
\om_n &+& g \left( \om_n^3 + 3\om_0 \om_n^2 + 3\om_0^2 \om_n +
\frac{6}{n} \left( \om_0 + \om_n \right)  \right) = 0 \;.
\eea

This system of equations has a trivial solution

\be
        \om_0 = \om_n = 0 \;,
\ee

corresponding  to the symmetric  phase with vanishing  VEV of the
field.  The non-trivial solutions cannot be obtained exactly.  If
we assume  that  there  is a critical  coupling  where  the field
starts to develop a VEV, then very close to this coupling the VEV
$\om_0$  should be small and we expand $\om_n$  as a power series
in $\om_0$.  This was already done in \cite{HKS92}, and we merely
quote the result (note the different normalization),

\bea
\om_n  &=& - \frac{6g}{n+6g}  \om_0 + \frac{gn}{n+6g}  \left(  1-
\left(\frac{n}{n+6g}   \right)^3  \right)  \om_0^3  +  O(\om_0^5)
\nonumber \\
&\equiv&  \alpha_n  (g) \om_0 + \beta_n (g) \om_0^3  + O(\om_0^5)
\;. \label{OMN}
\eea

Inserting  this  in  (\ref{ANSVAC})  one  obtains  $\om_0$  as  a
function of the coupling $g$ via

\be
1  +  g\sum_{n>0}  \frac{\alpha_n}{n}   +  g \, \om_0^2  \left(  1+
\sum_{n>0} \frac{\beta_n}{n} \right) = 0 \; .
\ee

As $\beta_n >0 $, this equation develops two real solutions  with
opposite sign if

\be
1+ g \sum_{n>0} \frac{\alpha_n}{n} \le 0  \; .
\label{SSBCOND}
\ee

The critical  coupling  $g_c$  is determined  if equality  holds.
Using the explicit form of $\alpha_n$ from (\ref{OMN}), one finds

\be
1 - g_c [\Psi (1+6g_c) + \gamma] = 0 \; ,
\label{GCRIT}
\ee

with a numerical value of the critical coupling

\be
g_c = 0.53070059\ldots \; .
\ee

which   in   terms   of   the   original   parameters   is  ({\it
cf.}~\cite{WX95})

\be
\la_c = 24 \pi g_c m^2 \simeq 40.0 m^2 \; .
\ee

Above  this  coupling,  the  VEV  $\om_0$  is  non-vanishing   and
acquires  one  of  two  possible  signs  so that  the  reflection
symmetry  is  spontaneously  broken.  Note  again  the  different
normalization of the coupling compared to \cite{HKS92},

\be
g_{\mbox{\tiny old}} = 6\,g_{\mbox{\tiny new}} \; .
\ee

\begin{figure}
\begin{center}
\hspace{1pt}\epsfysize=8.5cm \epsfbox{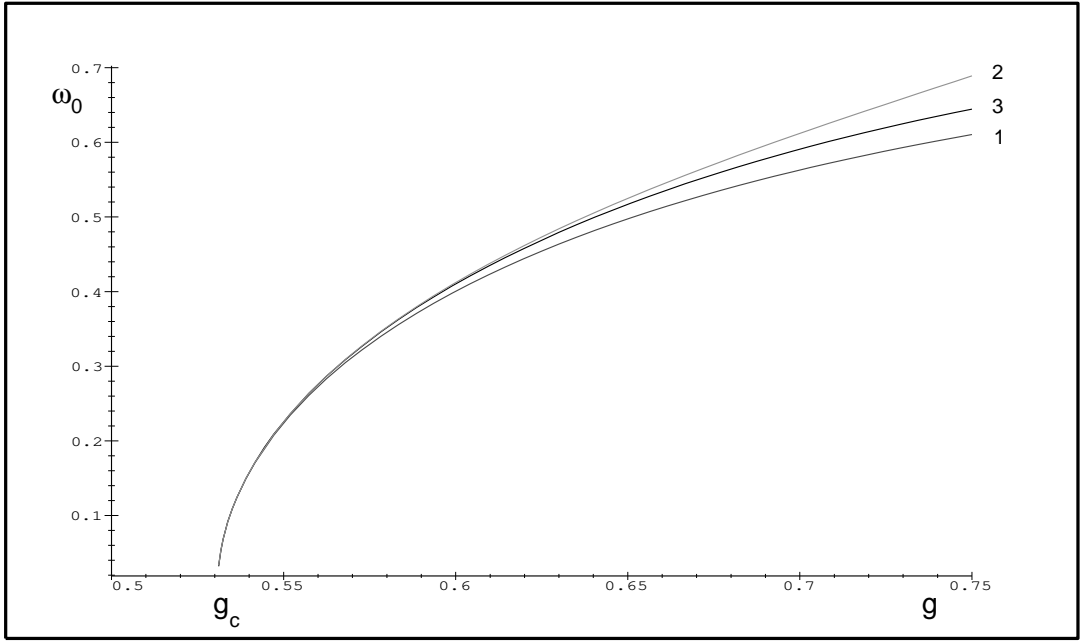}
\caption{\it Behaviour of the rescaled
VEV $\om_0$ close to the critical coupling $g_c$.
Graphs {\rm 1},{\rm 2} and {\rm 3} refer to the expansion
{\rm (\ref{OMN})} of $\om_n$
in $\om_0$ up to third, fifth and seventh order, respectively.}
\label{fig-omeg}
\end{center}
\end{figure}

The dependence  of the rescaled VEV $\om_0$ is plotted in Fig.\ref{fig-omeg}.
The critical exponent is 1/2, implying  mean-field  behaviour  as
expected  from the simple  form of the ansatz (\ref{MFA}).   Also
note the non-analyticity of $\om_0$ in the coupling at $g_c$.

\ind
{}From  the  one-particle   energies   (\ref{ENERGIES2})   one  can
calculate the mass-shift induced by the non-perturbative ZM $\om$
as given by the ansatz (\ref{MFA}),

\bea
\delta m^2 &=& 2P^\p_n P^\m_n - m^2 = 2\frac{\pi  n}{L} H_n - m^2
= m^2 \Bigg[ \frac{n}{2}  \left( \om_n^2 + 2\om_0 \om_n \right) +
\nonumber \\
&+& \frac{ng}{4}\left( \left( \om_0 +\om_n \right)^4-\om_0^4 +
\frac{12}{n}       \left(      \om_0+\om_n       \right)^2      +
4\om_n\sum_{k=1}^{\infty}{\frac{\om_k}{k}}+2\frac{\om_n^2}{n}
\right) \Bigg] \; .
\eea

Using the expression  (\ref{OMN})  for $\om_n$ this becomes (near
the critical coupling)

\bea
\delta  m^2  &=&  m^2  \om_0^2   \Bigg\{   3g  -  6g  \left[1  -
\frac{g}{1+6g/n} \bigg(\Psi(1+6g) + \gamma \bigg) \right] +
\nonumber \\
&+& 18g^2 \frac{n + 7g}{(n+ 6g)^2} \Bigg\} \; .
\eea

In contrast  to the  perturbative  result  (\ref{SHIFTRES}), this
expression  is {\it not} vanishing  in the continuum  limit  ($n,L
\to \infty$, $n/L$ finite),

\be
\delta m^2 \to  m^2  \om_0^2   \biggl\{   3g  -  6g  \Big[ 1  -
g \big(\Psi(1+6g) + \gamma \big) \Big] \biggr\} \; .
\ee

Due to (\ref{SSBCOND})  and(\ref{GCRIT})  the term  in the square
brackets is a small (negative)  quantity.  The mass shift induced
by $\om$ is thus positive. We want to emphasize that the ZM $\om$
has a non-trivial influence on the spectrum.





\vspace{1cm}

\section{The Tamm-Dancoff Approximation}

\vspace{1cm}

\subsection{General Remarks}

\vspace{1cm}

The  Tamm-Dancoff  approximation  (TDA)  was originally  designed
within  the  equal-time  formulation   of  quantum  field  theory
\cite{Tam45}. The core of the method was to enormously reduce the
particle  number  to  a  finite  (and  small)  one.  Due  to  the
complicated   many-body  structure  of  the  equal-time   vacuum,
however, the approximation failed to lead to quantitative results
and was abandoned thereafter.   Although it has been noted rather
early  that LF field  theory  might  be better  suited  for a TDA
\cite{tHo74}  , it was not until  recently  that people  began to
start systematic investigations  \cite{HPW90}.  The hope was (and
is), of course, that many of the problems of the original TDA can
be avoided due to the simplicity of the LF vacuum.  Specifically,
people tried to calculate bound states for 1+1 dimensional  field
theories and were able to obtain reasonable results \cite{HP91}.

\ind
However,  it became  clear in the meantime,  that there remain  a
number  of  open  problems  in connection  with  renormalisation,
especially  in more than 1+1 space-time  dimensions  \cite{Per94,
Wil94}.  The structure of counterterms  as well as their required
number is unclear.   Symmetries  like rotational  invariance  are
often  violated  \cite{Ji94}.    In  recent  publications  on  LF
$\phi^4$-theory  with  a TD  truncation  there  even  remained  a
logarithmic  UV divergence, the origin of which was rather unclear
\cite{PS94}.

\ind
In the following  we will try to shed some light on the issue
of  renormalisation,  in particular  on the  construction  of the
counterterms needed.  These could, in principle, be different for
different     particle    number    sectors    ("sector-dependent
renormalisation" \cite{HPW90}). For a super-renormalisable theory
like $\phi^4_{1+1}$,  which conventionally  is renormalisable  by
normal-ordering,  this would be a rather undesired  feature since
it would  complicate  the renormalisation  procedure  enormously.
Therefore,   in  the  following,   we  will   try  to  keep   the
renormalisation as simple as possible. To this end, we attempt to
incorporate the normal-ordering prescription into the TDA.  First
we need some definitions.  One- and two-particle states are given
by

\bea
\vert n \ket &=& a_n^\dagger \vert 0 \ket \\
\vert m,n\ket &=& a_m^\dagger a_n^\dagger \vert 0 \ket
\eea

with normalization

\bea
\bra m \vert n \ket &=& \delta_{mn} \\
\bra   k,l   \vert   m,n   \ket   &=&  \delta_{km}\delta_{ln}   +
\delta_{kn}\delta_{lm} \; .
\eea

The projection operators on the lowest particle number states are

\bea
P_0 &=& \vert 0 \ket \bra 0 \vert \\
P_1 &=& \sum_{n>0} \vert n \ket \bra n \vert \\
P_2 &=& \frac{1}{2}  \sum_{n,  m > 0} \vert  n,m\ket  \bra  m, n \vert \\
\nonumber         &\vdots&\\
P_N &=& \frac{1}{N!} \sum_{n_1,\ldots,n_N > 0}
\vert n_1,\ldots,n_N \ket  \bra n_1,\ldots,n_N  \vert
\; .
\eea

Most important will be the Tamm-Dancoff projector

\be
\Pn \equiv \sum_{\alpha=1}^{\mbox{
\scriptsize \sf N}} P_\alpha \; ,
\ee

which projects  onto the direct sum of all sectors  with particle
number  less  than or equal  to {\sf N}.  Before  we perform  any
detailed  calculation,  let  us  make a  few  remarks  about  the
relativistic invariance of the TDA \cite{Sch95}.

\ind
Let ${\cal  P}$ denote  the Poincar\'e  group.   In $d$ space-time
dimensions the number of Poincar\'e generators is

\be
\label{DIM-P-GROUP}
\mbox{dim} \, {\cal P} = \frac{d(d+1)}{2} \equiv D\; .
\ee

If the Poincar\'e  generators are $G_0, G_1, \ldots G_{D-1}$, the
TDA is relativistically invariant if

\be
[\Pn G_i \Pn , \Pn G_k \Pn] = \Pn [G_i , G_k ] \Pn \; .
\ee

It is easy to see, that this expression  can only hold if at most
one of the Poincar\'e generators fails to commute with $\Pn$.  So
we must have {\it e.g.}

\bea
[G_0, \Pn] &\not=& 0 \; , \\
\lbrack G_k , \Pn] &=& 0 \; , \; \;  k = 1, \ldots D-1 \; .
\label{INV}
\eea

The last identity implies that the $G_k$ conserve particle number
and  therefore   must  be  kinematical,   {\it  i.e.}~interaction
independent.   Thus, $G_0$ must be the only dynamical  generator.
Clearly, this can only happen if the dimension  of the Poincar\'e
group itself is small and if its stability  group (of kinematical
generators)  is large.   There  is only one single  case when the
requirements (\ref{INV}) are met, namely light-front quantisation
in $d$ = 1+1,  the case we are discussing  in this paper!   Here,
according   to  (\ref{DIM-P-GROUP}),   there   are   only   three
generators;  the dynamical  generator  is $G_0 \equiv  P^-$,  the
light-front  energy; the kinematical  generators are the momentum
$P^+$ and the (longitudinal)  boost $M^{+-}$.  We think that this
unique  feature  is one  of the  reasons,  why the TDA  (or  more
generally:   Fock space truncation  methods)  work so well for LF
field theories in $d$ = 1+1.

\ind
Encouraged  by this  observations  we continue  and  analyse  the
impact  of the  TDA on the  quantum  nature  of our scalar  field
theory.  This will be important for the issue of normal-ordering.
We imagine  that any operator  ${\cal  O}$ can be built  from the
elementary  Fock operators  $a_n$, $a_n^\dagger$,  no matter  how
complicated   its  form  may  be.   Thus,  we  {\it  define}  the
$N$-particle ($N$P) TDA by the replacement

\bea
a_n &\to& \Pn a_n \Pn \; , \label{NPLFTD1}\\
a_n^\dagger &\to& \Pn a_n^\dagger \Pn \label{NPLFTD2} \; .
\eea

In this way, however, one is mutilating  the quantum structure of
the theory.  This can be seen by calculating  the $N$PTDA  of the
elementary commutator

\be
[a_m , a_n^\dagger]_{\sfn} \equiv  [\Pn a_m
\Pn , \Pn a_n^\dagger \Pn] \; .
\ee

For   arbitrary   {\sf   N},   this   commutator   is  not  quite
straightforwardly  evaluated,  so let us briefly  go through  the
relevant steps. Firstly, we have

\be
[a_m, a_n^\dagger ]_{\sfn} = \delta_{mn} \Pn
+ \Pn [a_m , \Pn] a_n^\dagger \Pn + \Pn a_n^\dagger  [a_m , \Pn ]
\Pn \; , \label{TDCOMM}
\ee

so that we need

\be
[a_n , \Pn ] = (\Pnm - \Pn ) a_n = - P_N a_n \; ,
\ee

which can be found inductively. Using the additional identity

\be
\Pn P_N = P_N \; ,
\ee

expression (\ref{TDCOMM}) becomes

\be
[a_m  , a_n^\dagger  ]_{\sfn} = \delta_{mn}
\Pnm - P_N a_n^\dagger  a_m \Pn - \Pn a_n^\dagger  P_Na_m\Pn \; .
\ee

It is important  to note that the operator $a_n^\dagger  P_N a_m$
has non-vanishing  matrix elements  in the $N+1$-particle  sector
only.  The same is true for the operator $P_N a_n^\dagger a_m$ in
the $N$-particle sector. This leads to the final result

\be
[a_m  , a_n^\dagger  ]_{\sfn} = \delta_{mn}
\Pnm - P_N a_n^\dagger a_m P_N \; .
\label{TDCOMM2}
\ee

Taking matrix elements  of this expression  one readily sees that
the  correct  result  ($\delta_{mn}$)  for the commutator  within
$N$PTDA is reproduced only up to the ($N$-1)-particle sector. The
additional term on the {\it r.h.s.}~of (\ref{TDCOMM2})  is acting in the
$N$-particle  subspace  only.  In other words, matrix elements of
expressions   involving  the  elementary  commutator,  which  are
calculated  within  $N$PTDA,  should  not be trusted  beyond  the
($N$-1)-particle sector. This will be relevant for the problem of
normal-ordering to be discussed shortly.

\ind
To  be a little  bit  more  explicit,  we list  the  lowest  order
expressions for (\ref{TDCOMM})

\bea
[a_m , a_{n}^{\dagger}  ]_1 &=& \delta_{mn} \vert 0 \ket \bra 0 \vert
- \vert n \ket \bra m \vert \; , \\
\lbrack a_m , a_{n}^{\dagger} ]_2 &=& \delta_{mn} \Big[ \vert 0 \ket \bra 0
\vert + \sum_{l>0} \vert l \ket \bra l \vert \Big] - \sum_{l>0}
\vert l,n \ket \bra m,l \vert \; .
\eea

What   are   now   the   implications   of  all  that   for   the
renormalisation,  in particular mass renormalisation?   As can be
seen   from   (\ref{WICK}),   the  latter   is  encoded   in  the
normal-ordering prescription of the expression

\bea
\frac{1}{2L} \intl dx^\m \vi^2 (x) &=& \frac{1}{2L} \intl dx^\m :
\vi (x)^2 : + T  \nonumber \\
&=&  \sum_{n>0}  \frac{1}{2\pi  n} a_n^\dagger  a_n  + \sum_{n>0}
\frac{1}{4\pi n} [a_n, a_n^\dagger] \; .
\label{NORD}
\eea

Normal-ordering  thus  amounts  to splitting  off  the  divergent
tadpole  contribution  $T$, which, in the Fock space language, is
given  by an elementary  commutator  (or contraction).   Thus the
remarks  above,  leading  to  (\ref{TDCOMM2}),   apply.   Let  us
calculate the $N$PTDA of (\ref{NORD}),

\bea
\frac{1}{2L}   \intl   dx^\m   \vi^2   (x)  \stackrel{\mbox{\tiny
$N$PTDA}}{\simeq}    &=&   \sum_{n>0}   \frac{1}{2\pi   n}   \Pnm
a_n^\dagger  \Pnm a_n \Pnm + \Pnm T - \nonumber \\
&-& \sum_{n>0} \frac{1}{4\pi n} P_N a_n^\dagger a_n P_N \; .
\eea

The same  remarks  as for (\ref{TDCOMM2})  are in order.   Matrix
elements  of the last  expression  should  not be expected  to be
consistent beyond the ($N$-1)-particle  sector.  Furthermore, one
should  note that, as $\vi^2$  is a one-body  operator,  there is
only {\it one} commutator (or contraction)  involved in the above
normal-ordering.  For a $k$-body operator we therefore conjecture
that  its  renormalisation   will  be  correct  only  up  to  the
($N-k$)-particle  sector.  For  example,  in  order  to  get  the
renormalisation  of the two-body  operator  $\vi^4$ correct up to
the one-particle sector, a three-particle TDA will be needed.

\vspace{1cm}

\subsection{One-Particle Light-Front Tamm-Dancoff Approximation}

\vspace{1cm}

The     one-particle      LFTDA     is     defined     by     the
replacements\footnote{Usually,  the  LFTDA  refers  only  to  the
technique  of solving  bound state problems  by using  a few-body
ansatz for the bound-state wave function\cite{HPW90,HP91}. We use
the term LFTDA  in a wider context,  defined  essentially  by the
replacements (\ref{NPLFTD1}) and (\ref{NPLFTD2}).}

\bea
a_n &\simeq&  \Proj_1 a_n \Proj_1  = \vert 0 \ket \bra n \vert \\
a_n^\dagger  &\simeq&  \Proj_1 a_n^\dagger\Proj_1  = \vert n \ket
\bra 0 \vert \; .
\eea

{}From our general results we expect that within one-particle LFTDA
we will get a consistent renormalisation of the one-body operator
$\vi^2$ in the vacuum sector only.

\ind
To solve for the ZM $\om$ we make the following TD ansatz

\be
\om_{TD} = c_0 \vert 0 \ket \bra 0 \vert +  \sum_{n>0}  c_n \vert n
\ket \bra n \vert \; .
\label{OMTD1}
\ee

We similarly  expand the constraint  $\theta$ and the Hamiltonian
$H$

\bea
\theta   &\simeq&  \bar\theta_0  \vert  0  \ket  \bra  0  \vert  +
\sum_{n>0} \bar\theta_n \vert n \ket \bra n \vert \; , \\
H &\simeq&  \bar H_0 \vert 0 \ket \bra 0 \vert + \sum_{n>0}  \bar
H_n \vert n\ket \bra n \vert \; ,
\eea

where   the  bars  simply   indicate   the  distinction   of  the
coefficients   above  from  those  of  the  MF ansatz  (\ref{MFT})  and
(\ref{MFHAM}).   The coefficients  can be found  as functions  of
$c_0$,  $c_n$ upon inserting  the ansatz  (\ref{OMTD1})  into the
constraint (\ref{T1}) and the Hamiltonian (\ref{HAM}) yielding

\bea
\bar \theta_0   &=&  \left(m^2   +  \frac{\la}{3}   T  \right)  c_0  +
\frac{\la}{3!}   \left(c_0^3  +  \sum_{n>0}  \frac{c_n}{4\pi   n}
\right) \; , \label{T0} \\
\bar \theta_n  &=&  \left(  m^2 + \frac{\la}{12  \pi n} \right)  c_n +
\frac{\la}{3!} \left( c_n^3 + \frac{c_0}{4\pi n} \right) \; ,
\eea

and

\bea
\frac{\bar H_0}{2L} &=& \frac{1}{2} \left( m^2 + \frac{\la}{4}  T \right) c_0^2
+
\frac{\la}{4!} \left( c_0^4 + 2 \sum_{n>0} \frac{c_0c_n}{4\pi  n}
+ \sum_{n>0}  \frac{c_n^2}{4\pi  n} \right)  +
\frac{m^2}{2}  T + \frac{\la}{4!} T^2 , \label{H0} \\
\frac{\bar H_n}{2L}  &=& \frac{1}{2}  \left( m^2 + \frac{\la}{16 \pi n} \right)
c_n^2 + \frac{1}{2} \left( m^2 + \frac{\la}{12} T \right)
\frac{1}{4\pi n}
+ \frac{\la}{4!}  \left ( c_n^4 + \frac{c_0  c_n}{2\pi n} +
\frac{c_0^2}{4\pi n} \right) \!\!.
\eea

At a first look, these expressions for the coefficients appear to
be a disaster:  the infinities  in form of the tadpole $T$ do not
appear systematically,  the mass gets renormalised  in the vacuum
sector only, but differently  for the constraint $\theta$ and the
Hamiltonian  $H$.  Both  expressions  differ  from  the  standard
expression   $m^2  +  \la  T/2$.   There   is  also  a  divergent
contribution  from the $\vi^4$  term to $\bar  H_0$ which differs
from the usual $\la T^2 /8$ of (\ref{WICK}).  As stated above, we
do not believe the coefficients $\bar \theta_n$ and $\bar H_n$ to
be correct within one-particle  LFTDA.  They will be discussed in
the next subsection, when we go to higher order.

\ind
The  way  to remedy  the situation  (for  the coefficients  $\bar
\theta_0$ and $\bar H_0$) is the following.   We insist, firstly,
on the standard  mass renormalisation,  $m^2 + \la T/2$, however,
according  to our general discussion,  in the vacuum sector only.
Secondly, we do not assume that the coefficients  $c_0$ and $c_n$
are independent,  and use this freedom to {\it redefine} $c_n$ in
the following way

\bea
c_0 &\equiv& \om_0 \; , \\
c_n &\equiv& \om_0 + \om_n \; .
\label{REDEF1}
\eea

Inserting this into (\ref{T0}) and (\ref{H0}) one finds

\bea
\bar \theta_0  &=&  \left(  m^2  +  \frac{\la}{2}   T  \right)\om_0  +
\frac{\la}{3!}  \left( \om_0^3 + \sum_{n>0} \frac{\om_n}{4\pi  n}
\right) \; ,  \label{1PT0} \\
\bar H_0/2L &=& \frac{1}{2} \left( m^2 + \frac{\la}{2} T \right) \om_0^2 +
\frac{\la}{4!}   \left(   \om_0^4   +   4\sum_{n>0}   \frac{\om_0
\om_n}{4\pi  n}  + \sum_{n>0}  \frac{\om_n^2}{4\pi  n} \right)  +
\nonumber \\
&+& \frac{m^2}{2} T + \frac{\la}{4!} T^2 \; . \label{1PH0}
\eea

Remarkably, the simple redefinition (\ref{REDEF1}) has led to the
desired results.   The mass renormalisation  is standard  and the
same for $\theta_0$ and $H_0$.  The divergences  thus can be made
to  vanish  by  adding  the  counterterm  (\ref{COUNTER2}).  Both
equations (\ref{1PT0}) and (\ref{1PH0})  coincide with the lowest
order  results  from  the  mean-field  ansatz  (\ref{T0MF})   and
(\ref{E0MF})  (up to the constant  $\vi^4$-contribution  to $H_0$
given by $\la T^2/4!$).   Note that there are no two-body ($T^2$)
contributions  to the constraint.   This is obviously true to all
orders, so the renormalisation  of $\theta$  is slightly  simpler
than that of $H$, namely just mass renormalisation.

\ind
The coincidence  with the MF results  is not accidental.   If one
calculates  the lowest  order matrix  elements  of the MF
ansatz (\ref{MFA}), one finds

\bea
\om_0 = \bra 0 | \om_{MF} | 0 \ket &=& \bra 0 | \om_{TD} | 0 \ket = c_0
\; , \\
\om_0 + \om_n = \bra n | \om_{MF} | n \ket &=& \bra n | \om_{TD} | n
\ket = c_n \; .
\eea

Analogous   relations   hold  for  the  matrix  elements  of  the
constraint and the Hamiltonian,

\bea
\bra 0 | \theta | 0 \ket &=&  \theta_0 = \bar \theta_0 \; , \\
\bra 0 | H | 0 \ket &=& H_0 = \bar H_0 \; , \\
\bra  n  |  \theta  | n \ket  &=&  \theta_0  + \theta_n  = \bar
\theta_n  \;  , \\
\bra  n | H | n \ket  &=&  H_0 + H_n = \bar H_n \; .
\eea

Thus,  after  the  redefinition  (\ref{REDEF1}),  the  zero-  and
one-particle  matrix elements  of $\om$ calculated  within the MF
ansatz and TDA coincide.  As the renormalisation within MF ansatz
was  conventional  and straightforward,  it is not too surprising
that  the behaviour  of the redefined  TDA under  renormalisation
gets improved. This will be another guideline in the following.

\vspace{1cm}

\subsection{Two-Particle Light-Front Tamm-Dancoff Approximation}

\vspace{1cm}

If we now go one  step  further  and  include  also  two-particle
states via

\bea
a_n &\simeq&  \Proj_2  a_n \Proj_2  = \vert 0 \ket \bra n \vert +
\sum_{m>0} \vert m \ket \bra m,n \vert \\
a_n^\dagger  &\simeq& \Proj_2 a_n^\dagger  \Proj_2 = \vert n \ket
\bra 0 \vert + \sum_{m>0}  \vert n , m \ket \bra m \vert
\; ,
\eea

we should further improve our renormalisation program.  We expect
a consistent  renormalisation  of  $\vi^2$-contributions  in  the
vacuum- and one-particle sector, and of $\vi^4$-contributions  in
the vacuum sector. The extended ansatz for $\om$ becomes

\bea
\om_{TD} &=& c_0 | 0 \ket \bra 0 | + \sum_{n>0} c_n | n \ket \bra n
| + \nonumber \\
&+&  \sfrac{1}{2}  \sum_{m,n>0}  c_{mn} \, | m,n \ket  \bra  m+n | +
\sfrac{1}{2}  \sum_{m,n>0}  c_{mn}^* \, |  m+n  \ket  \bra  m,n  | +
\nonumber \\
&+& \sfrac{1}{4} \sum_{k,l,m,n>0}  \delta_{k+l,  m+n} \, c_{klmn}
\, | k,l \ket \bra m,n| \; .
\label{OMTD2}
\eea

It is now very plausible  (though  we cannot  prove  it a priori)
that a consistent renormalisation requires a redefinition also of
the coefficients $c_{mn}$, $c_{mn}^*$, and $c_{klmn}$.  To proceed
as  before,  we  would  need  the  matrix  elements
$\bar \theta_0$, $\bar \theta_n$, $\bar H_0$, and $\bar H_n$ with
all two-particle  contributions
in  order  to just  get  a consistent  renormalisation  up to the
one-particle sector.  This is very tedious and inefficient, as we
are {\it only}  interested  in the {\it divergent}  contributions
from  higher   order  terms  to  lower  order  matrix   elements.
Fortunately,  there is an alternative:  we simply demand that the
matrix elements of $\om_{TD}$ and $\om_{MF}$ coincide also in the
two-particle  sector.   This gives us the desired  redefinitions,
namely

\bea
c_{mn} &=& 2 \om_{mn} \; , \\
c_{mn}^* &=& 2 \om_{mn}^* \; , \\
c_{klmn}  &=& (\om_0 + \om_n + \om_n)  (\delta_{km}\delta_{ln}  +
\delta_{kn}\delta_{lm} ) + 4 \, \delta_{k+l, m+n}\, \om_{klmn} \; .
\eea

Thus,  essentially,   only  the  two-particle   matrix   elements
$c_{klmn}$ get redefined. Expression (\ref{OMTD2}) becomes

\bea
\om_{TD} &=& \om_0 | 0 \ket \bra 0 | + \sum_{n>0} (\om_0 + \om_n)
| n \ket \bra n | + \nonumber \\
&+& \sum_{m,n>0}  \om_{mn}  | m,n \ket \bra  m+n | + \sum_{m,n>0}
\om_{mn}^* | m+n \ket \bra m,n | + \nonumber \\
&+& \sfrac{1}{2}\sum_{m,n>0}  (\om_0 + \om_m + \om_n)
| m,n \ket \bra m,n | + \nonumber + \sum_{k,l,m,n>0} \delta_{k+l,
m+n} \, \om_{klmn} \, | k,l \ket \bra m,n| \; .
\eea

The diagonal two-particle term in the last line above contributes
to the equations determining the coefficients $\om_0$ and $\om_n$
and  crucially   alters   the  renormalisation   behaviour.    In
\cite{PS94}, it was noted that our MF ansatz amounts to including
two-particle  matrix elements, and it is just these terms that we
have now explicitly displayed.   Neglecting  all terms containing
$\om_{mn}$,  $\om_{mn}^*$,  and  $\om_{klmn}$,  which  are not of
interest within two-particle TDA, one finds for the constraint

\bea
\bar\theta_0  &=&  \left(  m^2  +  \frac{\la}{2}   T  \right)\om_0  +
\frac{\la}{3!}  \left( \om_0^3 + \sum_{n>0} \frac{\om_n}{4\pi  n}
\right) \; ,
\label{2PT0} \\
\bar\theta_n &=& \left( m^2 + \frac{\la}{2}  T \right) (\om_0 + \om_n)
+  \frac{\la}{3!}  \left[  (\om_0  + \om_n)^3  + 6 \frac{\om_0  +
\om_n}{4\pi  n} + \sum_{k>0}  \frac{\om_k}{4\pi  k} \right]  \; ,
\label{2PTN}
\eea

and for the Hamiltonian

\bea
\bar H_0/2L &=& \frac{1}{2} \left( m^2 + \frac{\la}{2} T \right) \om_0^2 +
\frac{\la}{4!}   \left[   \om_0^4   +   4\sum_{n>0}   \frac{\om_0
\om_n}{4\pi  n}  + \sum_{n>0}  \frac{\om_n^2}{4\pi  n} \right]  +
\nonumber \\ &+& \frac{m^2}{2} T + \frac{\la}{8} T^2 \; ,
\label{2PH0} \\
\bar H_n/2L &=& \frac{1}{2} \left( m^2 + \frac{\la}{2} T \right) (\om_0 +
\om_n)^2  +  \frac{1}{2}  \left(  m^2  + \frac{\la}{3}  T \right)
\frac{1}{2\pi n} + \nonumber \\
&+&  \frac{\la}{4!}  \Bigg[  (\om_0  + \om_n)^4  + \Big[12 (\om_0  +
\om_n)^2 + 2 \om_n^2 \Big]  \frac{1}{4\pi  n} + \nonumber \\
&& \quad\quad +\: 4(\om_0 + \om_n )
\sum_{k>0} \frac{\om_k}{4\pi  k} + \sum_{k>0} \frac{\om_k^2}{4\pi
k} \Bigg] + \nonumber \\
&+& \frac{m^2}{2} T + \frac{\la}{4!} T^2 \; .
\label{2PHN}
\eea

Several  remarks  are  in order.   As $\theta$  does  not contain
two-body  components,  the renormalisation  is correct  up to the
one-particle sector.  (\ref{2PT0}) and (\ref{2PTN}) thus coincide
with (\ref{T0MF}) and (\ref{TNMF}). The coefficient $\theta_0$ is
not even changed by including the two-particle  contributions  as
can  be  seen  by comparing  with  (\ref{1PT0}).   In the  vacuum
coefficient  $\bar  H_0$,  the mass renormalisation  (due  to the
$\vi^2$ contributions) {\it and} the vacuum energy $m^2 T/2 + \la
T^2  /8$  (with   the  $T^2$  contribution   stemming   from  the
$\vi^4$-term)  are  correct,  as  expected.   So  $\bar  H_0$  is
consistently renormalised.  In the one-particle coefficient $\bar
H_n$, which should  be compared  with (\ref{HNMF}), only the mass
renormalisation in the $\om$-sector is correct, as this is due to
one-body contributions  like $\om^2\vi^2$.   As anticipated, mass
renormalisation  and vacuum energy stemming from the $\vi^4$-term
differ from the correct values by numerical factors. To get these
correctly,  one  would  have  to  perform  a three-particle  TDA.
Presumably,  this  would  only  change  the  coefficients  of the
divergent  terms,  whereas  the coefficients  of the finite terms
would remain the same. This would then be analogous to the change
in $\bar  H_0$ by going  from  one-particle  TDA (\ref{1PH0})  to
two-particle TDA (\ref{2PH0}).

\ind
Summarizing,  we can say that,  in order  to obtain  a consistent
renormalisation  within a $N$-particle LFTD, one has to ({\it i}\/)
include contributions  from ($N+1$)-particle  matrix elements  by
({\it ii}\/) appropriately  redefining  the coefficients  in the TD
ansatz.  In this way, the Fock ansatz method (\ref{MFA})  and the
TDA become completely equivalent.

\vspace{1cm}

\section{Discussion and Conclusion}

\vspace{1cm}

In  this  paper  we  have  reanalysed  the  vacuum  structure  of
light-front $\phi^4_{1+1}$-theory by comparing different methods
of solving  for the constrained  zero mode of the field operator.
Within perturbation  theory,  the ZM induces  a second order mass
correction  which is vanishing  in the infinite volume limit.  We
believe  that to all orders  in perturbation  theory  the ZM only
induces  finite size effects,  although  we do not have a general
proof.

\ind
We  have  presented  two  non-perturbative  methods  to obtain  a
solution for the ZM.  An ansatz in terms of an increasing  number
of Fock operators,  which  we have truncated  after  the one-body
term, seems to be the most economic procedure.  With considerably
more efforts, exactly  the same results can be obtained  within a
light-front  Tamm-Dancoff  approximation,  if the renormalisation
procedure  is  properly  chosen.  By  doing  so  the  logarithmic
divergence of \cite{PS94},  stemming from an uncancelled  tadpole
term, does no longer appear\footnote{The  main difference between
the cited work and ours is that in the former the authors  do not
restrict  the  particle  number  for  intermediate  states.  This
results  in a divergence  structure  different  from ours, and it
seems to be rather difficult to find a systematic renormalisation
procedure in this case.}.

\ind
With either method  we find a non-vanishing  VEV $\phi_c$  of the
field if the coupling  $\la$ exceeds  a critical  value of $\la_c
\simeq 40 m^2$, implying spontaneous  breakdown of the reflection
symmetry $\phi \to - \phi$.  As the VEV changes continuously, the
associated  phase transition  is of second order,  which has been
rigorously  established  for the model  at hand\cite{SG73}.   The
order  parameter  $\phi_c$  shows  a square-root  behaviour  as a
function  of  the  coupling,  so  that  the  associated  critical
exponent is $\beta = 1/2$. The critical behaviour is thus of mean
field type, which is wrong, as the $\phi^4_{1+1}$ model is in the
universality  class  of the two-dimensional  Ising  model;  thus,
$\beta$  should be $1/8$.  It is difficult  to say, whether  this
shortcoming  can be removed if one extends our approximations  to
higher  orders.   This  question,   of  course, deserves  further
investigations.

\ind
Another  problem  we have  to face  is the absence  of any volume
dependence  of the  phase transition.   We have been working in a
finite  spatial  volume  of length  $2L$, the length  scale  $L$,
however, drops out of the equation (\ref{GCRIT})  determining the
critical  coupling.   On the other hand, there cannot  be a phase
transition  in a finite  volume  due to topological  fluctuations
(kinks  and  anti-kinks)  which  have  non-vanishing  statistical
weight   for  $L<\infty$.    We  have   not  incorporated   these
fluctuations  by our choice of (periodic) boundary conditions, so
it is perhaps not too astonishing  that we do not obtain a volume
dependence. Work in this direction is underway.


\begin{thebibliography}{100}

\bibitem{Dir49}
P.A.M.~Dirac, Rev.~Mod.~Phys. {\bf 21}, 392 (1949)

\bibitem{Pen60}
R.~Penrose, Ann.~Phys.~(N.Y.) {\bf 10}, 171 (1960)\\
R.~Penrose, Gen.~Relativ.~Gravit.~{\bf 12}, 225 (1980), reprinted
from 1962 \\
R.K.~Sachs,  J.~Math.~Phys.~{\bf  3}, 908 (1962), Phys.~Rev.~{\bf
128}, 2851 (1962)\\
H.~Bondi,  M.~van der Burg, A.~Metzner,  Proc.~Roy.~Soc.~(London)
{\bf A269}, 21 (1962)\\
A.~Komar, Phys.~Rev.~{\bf 134}, B1430 (1964)

\bibitem{FF65}
S.~Fubini, G.~Furlan, Physics {\bf 1}, 229 (1965)\\
R.~Dashen, M.~Gell-Mann, Phys.~Rev.~Lett.~{\bf 17}, 340 (1966)

\bibitem{BS67}
K.~Bardakci, G.~Segr\'e, Phys.~Rev.~{\bf 159}, 1263 (1967)\\
L.~Susskind, Phys.~Rev.~{\bf 165}, 1535 (1968)\\
J.~Jersak, J.~Stern, Nucl.~Phys.~{\bf B7}, 413 (1968)\\
K.~Bardakci, M.B.~Halpern, Phys.~Rev.~{\bf 176}, 1686 (1968)\\
H.~Leutwyler,  in:  Springer  Tracts in Modern Physics,  Vol.~50,
G.~H\"ohler, ed., Springer, 1969

\bibitem{KS70}
J.B.~Kogut, D.~Soper, Phys.~Rev.~{\bf D1}, 2901 (1970)

\bibitem{LKS70}
H.~Leutwyler,  J.R.~Klauder, L.~Streit, Nuovo Cim.~{\bf 66A}, 536
(1970)

\bibitem{NR71}
R.A.~Neville, F.~Rohrlich, Phys.~Rev.~{\bf D3}, 1692 (1971)

\bibitem{CM69}
S.-J.~Chang, S.-K.~Ma Phys.~Rev.~{\bf 180}, 1506 (1969)

\bibitem{AFF73}
V.~De Alfaro, S.~Fubini, G.~Furlan, C.~Rossetti, Currents in Hadron Physics,
North Holland, Amsterdam (1993)\\
J.~Kogut, L.~Susskind, Phys.~Rep.~{\bf 8C}, 75 (1973)

\bibitem{Wei66}
S.~Weinberg, Phys.~Rev.~{\bf 150}, 1313 (1966)

\bibitem{Col66}
S.~Coleman, J.~Math.~Phys.~{\bf 7}, 787 (1966)

\bibitem{MY76}
T.~Maskawa, K.~Yamawaki, Progr.~Theor.~Phys.~{\bf 56}, 270 (1976)

\bibitem{FNP81}
V.A.~Franke,        Yu.~V.~Novozhilov,         E.V.~Prokhvatilov,
Lett.~Math.~Phys.~{\bf 5}, 239, 437 (1981)\\
V.A.~Franke, E.V.~Prokhvatilov, Sov.~J.~Nucl.~Phys.~{\bf 47}, 559
(1988); {\bf 49}, 688 (1989)

\bibitem{Wit89}
R.S.~Wittman,  in:  {\sl  Nuclear  and  Particle  Physics  on  the
Light-Cone},   M.B.~Johnson,    L.S.~Kisslinger,    eds.,   World
Scientific, Singapore 1989

\bibitem{HKW91}
T.~Heinzl,  S.~Krusche,  E.~Werner,  Phys.~Lett.~{\bf  B256},  55
(1991)\\
T.~Heinzl,  S.~Krusche,  E.~Werner,  Phys.~Lett.~{\bf  B275}, 410
(1992)

\bibitem{HKW91b}
T.~Heinzl,  S.~Krusche,  E.~Werner,  Phys.~Lett.~{\bf  B272},  54
(1991)

\bibitem{HWZ92}
T.~Heinzl, S.~Krusche,  E.~Werner, B.~Zellermann,  {\sl Zero Mode
Corrections  in Perturbative  Light  Cone Scalar  Field  Theory},
Regensburg preprint TPR 92-17, 1992 (unpublished)

\bibitem{HKS92}
T.~Heinzl, S.~Krusche,  S.~Simb\"urger,  E.~Werner, Z.~Phys.~{\bf
C 56}, 415 (1992)

\bibitem{MR92}
G.~McCartor, D.G.~Robertson, Z.~Phys.~{\bf C53}, 679 (1992)

\bibitem{Hor92}
K.~Hornbostel, Phys.~Rev.~{\bf D45}, 3781 (1992)

\bibitem{Rob93}
D.G.~Robertson, Phys.~Rev.~{\bf D47}, 2549 (1993)

\bibitem{BJS93}
R.W.~Brown,      J.W.~Jun,     S.M.~Shvartsman,      C.C.~Taylor,
Phys.~Rev.~{\bf D48}, 5873 (1993)

\bibitem{BPS93}
S.S.~Pinsky, B.~van de Sande, C.M.~Bender, Phys.~Rev.~{\bf  D48},
816 (1993)

\bibitem{PS94}
S.S.~Pinsky, B.~van de Sande, Phys.~Rev.~{\bf D49}, 2001 (1994)

\bibitem{HPS95}
S.S.~Pinsky,  B.~van de Sande, J.R.~Hiller, Phys.~Rev.~{\bf D51},
726 (1995)

\bibitem{KPP94}
A.C.~Kalloniatis, H.-C.~Pauli, Z.~Phys.~{\bf C63}, 161 (1994)\\
A.C.~Kalloniatis,   D.G.~Robertson,  Phys.~Rev.~{\bf  D50},  5262
(1994)\\
A.C.~Kalloniatis, H.-C.~Pauli, S.S.~Pinsky, Phys.~Rev.~{\bf
D50}, 6633 (1994)\\
A.C.~Kalloniatis,  H.-C.~Pauli, S.S.~Pinsky, {\sl Towards Solving
QCD  - The Transverse  Zero  Modes  in Light-Cone  Quantization},
preprint MPI-H-V3-1995, hep-th/9509020\\
A.C.~Kalloniatis,  {\sl On Zero Modes and the Vacuum Problem  - A
Study  of Scalar  Adjoint  Matter  in Two-Dimensional  Yang-Mills
Theory  via  Light-Cone-Quantisation},  preprint  MPI-H-V29-1995,
hep-th/9509027

\bibitem{Mae94}
M.~Maeno, Phys.~Lett.~{\bf B320}, 83 (1994)\\
M.~Tachibana,  {\sl Quantum Mechanics  of the Dynamical Zero Mode
in  $QCD_{1+1}$  on  the  Light-Cone},   preprint  KOBE-TH-95-01,
hep-th/9504026\\
Y.~Kim, S.~Tsujimaru, K.~Yamawaki, Phys.~Rev.~Lett {\bf 74}, 4771 (1995)

\bibitem{WX95}
X.~Xu,  H.J.~Weber,  {\sl Loop Expansion  in Light-Cone  $\phi^4$
Field Theory}, hep-ph/9507450

\bibitem{Dir50}
P.A.M.~Dirac, Canad.~J.~Math.~{\bf 2}, 129 (1950)\\
P.A.M.~Dirac,  {\sl Lectures on Quantum Mechanics}, Benjamin, New
York, 1964

\bibitem{HRT76}
A.~Hanson, T.~Regge, C.~Teitelboim,  {\sl Constrained Hamiltonian
Systems}, Academia Nazionale dei Lincei, Rome, 1976

\bibitem{BM73}
L.~Banyai, L.~Mezinescu, Phys.~Rev.~{\bf D8}, 417 (1973)

\bibitem{Ida76}
M.~Ida, Nuovo Cim.~Lett.~{\bf 15}, 269 (1976)\\
M.~Huszar, J.~Phys.~{\bf A9}, 1359 (1976)\\
P.~Senjanovic, Ann.~Phys.~{\bf 100}, 227 (1976)

\bibitem{FJ88}
L.~Faddeev, R.~Jackiw, Phys.~Rev.~Lett.~{\bf 60},1692 (1988)

\bibitem{Gar71}
C.S.~Gardner, J.~Math.~Phys.~{\bf 12}, 1548 (1971)\\
E.~Witten, Comm.~Math.~Phys.~{\bf 92}, 455 (1984)\\
M.J.~Bergvelt,   E.A.~de  Kerf,  CAP-NSERC  Summer  Institute  in
Theoretical  Physics,  Edmonton,  Alberta,  1987,  Vol.~I,  World
Scientific, Singapore, 1988

\bibitem{HZ93a}
A.~Harindranath, W.-M.~Zhang, Phys.~Rev.~{\bf D48}, 4868 (1993)

\bibitem{HW94}
T.~Heinzl, E.~Werner, Z.~Phys. {\bf C62}, 521 (1994)\\
T.~Heinzl,  in:  Proceedings  of the Workshop  {\sl Quantum Field
Theoretical   Aspects   of  High   Energy   Physics},   B.~Geyer,
E.-M.~Ilgenfritz, eds., CHS/University of Leipzig, 1993

\bibitem{CN94}
C.~Cronstr\o m, M.~Noga, Nucl.~Phys. {\bf B428}, 449 (1994)


\bibitem{AS70}
M.~Abramowitz,  I.A.~Stegun,  eds., {\sl Handbook of Mathematical
Functions}, Dover, New York, 1970

\bibitem{Tam45}
I.~Tamm, J.~Phys.~(Moscow) {\bf 9}, 449 (1945)\\
S.M.~Dancoff, Phys.~Rev.~{\bf 78}, 382 (1950)

\bibitem{tHo74}
G.~`t Hooft, Nucl.~Phys.~{\bf B75}, 461 (1974)\\
H.~Bergknoff, Nucl.~Phys.~{\bf B122}, 215 (1977)

\bibitem{HPW90}
A.~Harindranath, R.J.~Perry, K.G.~Wilson, Phys.~Rev.~Lett.~{\bf  65},
2959 (1990)

\bibitem{HP91}
A.~Harindranath, R.J.~Perry, Phys.~Rev.~{\bf D43},4051 (1991)\\
A.~Harindranath, R.J.~Perry, J.~Shigemitsu, Phys.~Rev.~{\bf D46},
4580 (1992)\\
Y.~Mo, R.J.~Perry, J.~Comp.~Phys.~{\bf 108}, 159 (1993)\\
K.~Harada, T.~Sugihara, M.~Taniguchi,  M.~Yahiro, Phys.~Rev.~{\bf
D49}, 4226 (1994)

\bibitem{Per94}
R.J.~Perry, Ann.~Phys.~{\bf 232}, 116 (1994)

\bibitem{Wil94}
K.G.~Wilson {\it et al.}, Phys.~Rev.~{\bf D49}, 6720 (1994)

\bibitem{Ji94}
C.-R.~Ji,  in:   {\sl  Theory  of Hadrons  and Light-Front  QCD},
S.~Glazek, ed., World Scientific, Singapore 1995,

\bibitem{Sch95}
M.~Schmidt,    talk   presented    at   the   Regensburg/Erlangen
``Graduiertenkolleg'' meeting, Waischenfeld, Germany, May 1995

\bibitem{SG73}
B.~Simon, R.~B.~Griffiths, Comm.~Math.~Phys.~{\bf 33}, 145 (1973)

\end{thebibliography}
\end{document}